\documentclass[12pt]{iopart}
\usepackage{hyperref}
\usepackage{verbatim}
\usepackage{latexsym}
\usepackage{soul}
\usepackage{graphicx}

\newcommand{\be}{\begin{eqnarray} \begin{aligned}}
\newcommand{\ee}{\end{aligned} \end{eqnarray} }

\usepackage{amsfonts}

\newcommand{\qed}{\nobreak \ifvmode \relax \else
      \ifdim\lastskip<1.5em \hskip-\lastskip
      \hskip1.5em plus0em minus0.5em \fi \nobreak
      \vrule height0.75em width0.5em depth0.25em\fi}

\newcommand{\beq}{\begin{eqnarray}}
\newcommand{\eeq}{\end{eqnarray}}

\begin{document}

\title{Weakly bound states in heterogeneous waveguides}
\author{Paolo Amore\footnote{paolo.amore@gmail.com}}
\address{Facultad de Ciencias, CUICBAS, Universidad de Colima,\\
Bernal D\'{i}az del Castillo 340, Colima, Colima, Mexico}
\author{Francisco M. Fern\'andez\footnote{fernande@quimica.unlp.edu.ar}}
\address{INIFTA (UNLP, CCT La Plata-CONICET), Division Qu\'imica Te\'orica, \\
Blvd. 113 S/N, Sucursal 4, Casilla de Correo 16, 1900 La Plata, Argentina}
\author{Christoph P. Hofmann\footnote{christoph.peter.hofmann@gmail.com}}
\address{Facultad de Ciencias, CUICBAS, Universidad de Colima, \\
Bernal D\'{\i}az del Castillo 340, Colima, Colima, Mexico}

\begin{abstract}
We study the spectrum of the Helmholtz equation in a two-dimensional infinite waveguide,
containing a weak heterogeneity localized at an internal point, and obeying Dirichlet
boundary conditions at its border. We prove that, when the heterogeneity corresponds to a locally
denser material, the lowest eigenvalue of the spectrum falls below the continuum threshold
and a bound state appears, localized at the heterogeneity. We devise a rigorous perturbation
scheme and derive the exact expression for the energy to third order in the
heterogeneity.
\end{abstract}
\maketitle



\section{Introduction}
\label{sec:intro}

The existence of ``trapped modes" in open geometries was first
proved by Ursell \cite{Ursell51,Ursell52} studying beach waves in
semi-infinite canals. Much later, Schult and collaborators
\cite{Schult89}, showed that the ground state of an electron
constrained to move in an infinite symmetric cross, but otherwise
free, is localized in the central region of the cross and its
energy falls below the continuum threshold (recently Amore et al.
\cite{Amore12a} have also studied the more general case of
asymmetric crosses and asymmetric T- and L-shaped waveguides).
Since Ref.~\cite{Schult89}, a large class of open geometries and of physical
problems have been studied, observing the emergence of one or more
localized modes. In particular, Exner and Seba \cite{Exner89} have
proved that the existence of a bound state for an electron
confined to a planar waveguide, with curvature decaying at
infinity and obeying Dirichlet boundary conditions at the border;
Goldstone and Jaffe \cite{Goldstone92} have proved that an
electron confined to an infinite tube of constant cross section,
in two or more dimensions, has always a bound state, when the tube
is not perfectly straight. The effect of bound states in infinite
non--straight waveguides has been studied in
\cite{Carini92,Carini93,Carini97,Wu91,Wu93,Bittner13}. It is worth
mentioning a recent pedagogical article by Londergan and
Murdock~\cite{Londergan12}, that illustrates different numerical
methods for the solutions of confined systems, in particular
two-dimensional waveguides.

Bulla and collaborators \cite{Simon97} have obtained the expression for the fundamental energy of an infinite two-dimensional
waveguide, where the upper border is slightly deformed, to second order in the parameter controlling the deformation. Their result proves that
a bound state is always present, whenever the deformation corresponds to a local enlargement of the waveguide. Exner and
Vugalter \cite{Exner97} have then considered the interesting question of what happens when the leading expression of Bulla et al.
vanishes, and whether a bound state can still exist. This situation corresponds to the case where the local deformation of the waveguide
contains both a local enlargement and a local shrinking  of the waveguide, which preserve the total area.
They showed that the problem presents a critical behavior, related to the size of the deformation, where the bound state can be lost
below a critical size.

A different but related problem, concerns the existence of bound states in open geometries, that are not necessarily deformed, but contain
local heterogeneities. For the solvable case of an infinite, straight, two-dimensional waveguide, containing a segment of different density,
Wang \cite{Wang14} has proved the existence of bound states. Moreover, at least in two dimensions, the problem of studying the
spectrum of a deformed waveguide can be reduced to the problem of studying the spectrum of a heterogeneous straight waveguide,
by using a suitable conformal map which transforms one domain into the other. In a similar way one could study the spectrum of
a waveguide both deformed and heterogeneous.

The purpose of the present work is to establish rigorously the
conditions for the existence of bound states on general planar
weakly deformed and heterogeneous waveguides, applying a
perturbation method and calculating the {\sl exact} general
expression for the energy of the fundamental mode to third order
in the parameter controlling the perturbation. When our formula is
applied to the study of homogeneous and weakly deformed waveguides
the "density" appearing  in our expression is directly related to
the conformal map used to transform the deformed waveguide into a
straight one (for an example of use of conformal maps to the
infinite waveguide see \cite{Bittner13, Sadurni10}). In this case
our second order formula should be equivalent to the formula of
\cite{Simon97} for a specific class of maps. Our results can also
be used in more than two dimensions, to describe straight
heterogeneous waveguides (conformal maps are limited to two
dimensions). Finally, we stress that the perturbation scheme used
in the present paper, could be applied to include higher order
perturbative contributions in a systematic way.

The paper is organized as follows: in Section \ref{sec:PT}  we describe the perturbation method used in the calculation
and derive the exact expressions for the energy of the ground state up to third order in the perturbative parameter;
in Section \ref{sec:VM} we reproduce the second order expression for the energy of the ground state using the variational method;
in Section \ref{sec:example} we discuss a simple solvable example of an infinite, straight, weakly heterogeneous waveguide and
reproduce the exact results to third order, using our formulas; finally, in Section \ref{sec:conclusion} we present our conclusions.

\section{Perturbation theory}
\label{sec:PT}

Following \cite{Amore11, Amore13a} we consider the Helmholtz equation for an inhomogeneous medium,
\begin{eqnarray}
\left( - \Delta \right) \Phi_n({\bf x}) = E_n \Sigma({\bf x}) \Phi_n({\bf x}) \, ,
\label{eq:Helmholtz_ht}
\end{eqnarray}
where ${\bf x} \in \Omega_d$, and $\Omega_d$ is a region of
$\Re^d$. This region may either be finite or infinite, and the
spectrum of the eigenvalues $E_n$ can either be discrete,
continuum or both. In particular, several aspects of the
behavior of the spectrum  of eq.~(\ref{eq:Helmholtz_ht}), in one
or more dimensions, have been studied by one of the present
authors. The aspects studied earlier include the description of
non--perturbative methods for the calculation of the lowest part
of the spectrum~\cite{ Amore11,Amore15,Amore10a}, of a
perturbation method for the calculation of the eigenvalues of
two-dimensional domains obtained from a small deformation of a
reference (solvable) domain~\cite{Amore10b}, the derivation of
spectral zeta functions associated to heterogeneous systems in one
and two dimensions~\cite{Amore12b} and the sum rules of
heterogeneous domains in one or more dimensions, for different
boundary conditions~\cite{Amore13a,Amore14a,Amore13b}. 

Observe that the density of the medium is a positive definite function,
$\Sigma(\bf{x}) >  0$ for ${\bf x} \in \Omega_d$.

If we introduce the functions
\begin{eqnarray}
\xi_n({\bf x}) \equiv \sqrt{\Sigma({\bf x})} \Phi_n({\bf x})\, ,
\end{eqnarray}
we see that the Helmholtz equation becomes
\begin{eqnarray}
\frac{1}{\sqrt{\Sigma}} \left( - \Delta \right)\frac{1}{\sqrt{\Sigma}} \xi_n({\bf x})  = E_n \xi_n({\bf x}) \, ,
\label{eq:hermite}
\end{eqnarray}
and the $\xi_n$ are eigenfunctions of the operator $\hat{O} \equiv \frac{1}{\sqrt{\Sigma}} \left( - \Delta \right)\frac{1}{\sqrt{\Sigma}}$.
Notice that the two equations are {\sl isospectral}.
It is a matter of convenience to work with one equation or the other~\footnote{In the following we will work
with the form of eq.~(\ref{eq:hermite}), which has the advantage of using manifestly Hermitian operators;
the approach using eq.~(\ref{eq:Helmholtz_ht}) is described in \ref{appb}.}.

To solve the problem perturbatively we assume
\begin{equation}
\Sigma({\bf x}) = 1 + \eta \sigma({\bf x}) \, ,
\end{equation}
where $|\eta| \ll 1$ is a dummy perturbation parameter that one sets to unity after the calculation
and
\begin{eqnarray}
E_n &=& \sum_{j=0}^\infty \eta^j E_n^{(j)} , \\
\xi_n({\bf x}) &=& \sum_{j=0}^\infty \eta^j \xi_n^{(j)}({\bf x}) \, .
\end{eqnarray}

We work with the operator
\begin{eqnarray}
\hat{H} \equiv \frac{1}{\sqrt{\Sigma}} \left( - \Delta \right)\frac{1}{\sqrt{\Sigma}} \label{eq_H}
\end{eqnarray}
on the strip defined by $(x,y) \in \left\{ - \infty < x < \infty ,  -b/2 \leq y \leq b/2 \right\}$. Dirichlet boundary conditions at $y= \pm b/2$ are
assumed. We also assume that $\sigma(x,y)$ corresponds to a localized inhomogeneity, i.e.,
\begin{equation}
\lim_{|x| \rightarrow \infty} \sigma(x,y) = 0  \, .
\end{equation}

For a straight and homogeneous waveguide the spectrum is continuous and therefore if one applies perturbation
theory to the general problem of an inhomogeneous waveguide using the straight and homogeneous waveguide
as the unperturbed problem, one inevitably encounters infrared divergences, due to the contribution of the states of
very soft momentum, just above the fundamental mode. To avoid the emergence of such divergences, in the different
context of quantum mechanics and quantum field theory, Gat and Rosenstein~\cite{Gat93} have devised a perturbation scheme
which uses a regularized free theory and obtained explicit results for the bound state of weak short range potentials (thus reproducing a result first obtained by Simon \cite{Simon76}) and for the mass of the bound state meson in the Thirring model.

Applying the ideas of Ref.~\cite{Gat93} to our case, we  modify the original operator in a way that it can support a bound state:
\begin{eqnarray}
\hat{H} \rightarrow \frac{1}{\sqrt{\Sigma}} \left( - \Delta - 2 \beta \delta(x) \right)\frac{1}{\sqrt{\Sigma}} \, ,
\end{eqnarray}
where $\beta>0$.  We use as unperturbed operator $\hat{H}_0 \equiv - \Delta - 2 \beta \delta(x)$,
and we will let $\beta \rightarrow 0$ at the end of the calculation.

The basis set of eigenfunctions of $\hat{H}_0$ is given by
\begin{eqnarray}
\Psi_{p,n}(x,y) = \psi_n(y)  \otimes \left\{
\begin{array}{ccc}
\phi_o(x) &,&  {\rm ground \ state} \, , \nonumber \\
\phi_p^{(e)}(x) &,& {\rm even} \, ,  \nonumber \\
\phi_p^{(o)}(x) &,& {\rm odd} \, ,
\end{array}
\right .
\end{eqnarray}
where the longitudinal wave functions are
\begin{eqnarray}
\phi_0(x) &=& \sqrt{\beta} e^{-\beta |x|} \, , \nonumber \\
\phi_p^{(e)}(x)  &=& \frac{\sqrt{2}}{\sqrt{p^2+ \beta^2}} \left[ p \cos (p x)-\beta \sin(p|x|) \right] \, , \nonumber \\
\phi_p^{(o)}(x)  &=& \sqrt{2}  \sin(p x) \, ,
\end{eqnarray}
and the transverse wave functions are
\begin{eqnarray}
\psi_n(y) &=& \sqrt{\frac{2}{b}} \sin \left[\frac{n \pi}{b} (y+b/2)\right] \, .
\end{eqnarray}

The corresponding eigenvalues are
\begin{eqnarray}
\epsilon_{0,n} &=& -\beta^2 +\frac{n^2\pi^2}{b^2} \, , \nonumber  \\
\epsilon^{(e)}_{p,n} &=& \epsilon^{(o)}_{p,n} = p^2 +\frac{n^2\pi^2}{b^2} \, .
\end{eqnarray}

It is convenient to use Dirac notation
\begin{eqnarray}
\phi_0(x) \rightarrow |0\rangle  \ \ ; \ \
\phi_p^{(e)}(x) \rightarrow |p^{(e)}\rangle   \ \ ; \ \
\phi_p^{(o)}(x) \rightarrow |p^{(o)}\rangle   \ \ ; \ \
\psi_n(y) \rightarrow |n\rangle ,
\end{eqnarray}
and
\begin{eqnarray}
\Psi_{p,n}(x,y) &\rightarrow& \left\{
\begin{array}{c}
|0,n\rangle \, , \\
|p^{(e)},n\rangle \, , \\
|p^{(o)},n\rangle \, . \\
\end{array}  \right.
\end{eqnarray}

The completeness of the longitudinal basis reads
\begin{eqnarray}
|0\rangle \langle 0 | + \int_0^\infty \frac{dp}{2\pi} \left[ |p^{(e)}\rangle \langle p^{(e)}|+|p^{(o)}\rangle \langle p^{(o)}|\right] = \hat{1} \, ,
\end{eqnarray}
while the completeness of the transverse basis is
\begin{eqnarray}
\sum_{n=1}^\infty |n\rangle \langle n | = \hat{1} \, .
\end{eqnarray}

We express the operator $\hat{H}$ of eq.~(\ref{eq_H}) in terms of the "unperturbed" operator $\hat{H}_0$,
expanding it for small inhomogeneities ($\Sigma = 1 + \sigma$, $|\sigma |\ll 1$):
\begin{eqnarray}
\hat{H} &=& \hat{H}_0  - \frac{1}{2} \left[ \hat{H}_0  \sigma + \sigma \hat{H}_0  \right]
+ \frac{1}{8} \left[ 2 \sigma \hat{H}_0 \sigma  + 3 \sigma^2 \hat{H}_0  + 3 \hat{H}_0 \sigma^2 \right]
\nonumber \\
&-& \left\{ \frac{3}{16} \left[ \sigma^2 \hat{H}_0 \sigma + \sigma \hat{H}_0  \sigma^2 \right]
+ \frac{5}{16} \left[ \sigma^3 \hat{H}_0  +  \hat{H}_0  \sigma^3 \right] \right\} + \dots \nonumber \\
&\equiv& \hat{H}_0 + V^{(1)}+  V^{(2)} + V^{(3)} + \dots
\end{eqnarray}

Now we use the standard Rayleigh-Schr\"odinger  perturbation
theory (RSPT) with the basis above. Notice that, in contrast to
the usual applications of RSPT, here the ``perturbation" contains
contributions of different orders in $\sigma$, and therefore
contributions of a given order in $\sigma$ can originate from
different orders of the perturbative expansion (see
ref.~\cite{Amore10b}).

\subsection{First order}

To first order we have
\begin{eqnarray}
E_0^{(1)} &=& \langle 0,1 |  V^{(1)} | 0,1 \rangle \nonumber \\
&=& - \frac{\beta}{2} \int_{-\infty}^{\infty} dx \int_{-b/2}^{b/2} dy \ e^{-\beta |x|} \psi_1(y)
\left[\hat{H}_0  \sigma + \sigma \hat{H}_0 \right] e^{-\beta |x|} \psi_1(y) \nonumber \\
&=&  - \beta \epsilon_{0,1} \int_{-\infty}^{\infty} dx \int_{-b/2}^{b/2} dy \ e^{-\beta |x|} \psi_1(y)
\sigma(x,y) e^{-\beta |x|} \psi_1(y) \, .
\end{eqnarray}

The correct physics is recovered after taking the limit $\beta \rightarrow 0$, for which one has
\begin{eqnarray}
\lim_{\beta\rightarrow 0} E_0^{(1)} = 0 \, .
\end{eqnarray}

\subsection{Second order}

We introduce the operator
\begin{eqnarray}
\hat{\Omega} \equiv \left[{\sum_{n=2}^\infty}  \frac{1}{\epsilon_{0,n} - \epsilon_{0,1}}  | 0,n \rangle \langle 0,n |
+ {\sum_{n=1}^\infty} \int_0^\infty \frac{dp}{2\pi} \frac{1}{\epsilon_{p,n} - \epsilon_{0,1}} | p,n \rangle \langle p,n | \right] \, .
\end{eqnarray}

The second order correction is simply
\begin{eqnarray}
E_0^{(2)} &=& \langle 0,1 |  V^{(2)} | 0,1 \rangle - \langle 0,1 |  V^{(1)} \hat{\Omega} V^{(1)} | 0,1\rangle \, .
\end{eqnarray}

Using the explicit expressions for $ V^{(1)}$ and $ V^{(2)}$ and simplifying, we obtain
\begin{eqnarray}
E_0^{(2)} = \epsilon_{0,1} \langle 0,1 | \sigma | 0,1\rangle^2 - \epsilon_{0,1}^2 \langle 0,1 |\sigma \hat{\Omega} \sigma  | 0,1\rangle \, ,
\end{eqnarray}
where the first contribution vanishes for $\beta \rightarrow 0$.

Using the explicit expressions for $G({\bf x},{\bf x}') = \langle {\bf x} | \hat{\Omega} | {\bf x}'\rangle$ given in the \ref{appa},
we write the second contribution
\begin{eqnarray}
\lim_{\beta \rightarrow 0^+} E_0^{(2)} &=&  -  \lim_{\beta \rightarrow 0^+}  \epsilon_{0,1}^2 \langle 0,1 |  \sigma \hat{\Omega} \sigma | 0,1\rangle   \nonumber \\
&=&  -  \lim_{\beta \rightarrow 0^+}  \epsilon_{0,1}^2  \int_{-\infty}^{\infty} dx \int_{-b/2}^{b/2} dy \int_{-\infty}^{\infty} dx' \int_{-b/2}^{b/2} dy'
\sigma(x,y) \sigma(x',y') \nonumber \\
& & \times  G(\mathbf{x},\mathbf{x}') \phi_0(x)\phi_0(x') \psi_1(y) \psi_1(y') \nonumber \\
&=& - \frac{\pi^4}{b^6}  \left[ \int_{-\infty}^{\infty} dx \int_{-b/2}^{b/2} dy \ \sigma(x,y)  \cos ^2\left(\frac{\pi  y}{b}\right)    \right]^2 \, .
\end{eqnarray}


\subsection{Third order}

The third order correction reads
\begin{eqnarray}
E_0^{(3)} &=& \langle 0,1 |  V^{(3)} | 0,1 \rangle - \langle 0,1 |  V^{(1)} \hat{\Omega}  V^{(2)} | 0,1 \rangle -
\langle 0,1 |  V^{(2)} \hat{\Omega}  V^{(1)} | 0,1 \rangle  \nonumber \\
&+& \langle 0,1 |  V^{(1)} \hat{\Omega}  V^{(1)}  \hat{\Omega}  V^{(1)} | 0,1  \rangle
- \langle 0,1 |  V^{(1)} | 0,1 \rangle  \langle 0,1 |  V^{(1)} \hat{\Lambda} V^{(1)}  | 0,1\rangle \, ,
\end{eqnarray}
where we have introduced the operator
\begin{eqnarray}
\hat{\Lambda} \equiv \hat{\Omega}^2 = \left[{\sum_{n=2}^\infty}  \frac{1}{(\epsilon_{0,1} - \epsilon_{0,n})^2}  | 0,n \rangle \langle 0,n |
+ {\sum_{n=1}^\infty} \int_0^\infty \frac{dp}{2\pi} \frac{1}{(\epsilon_{0,1} - \epsilon_{p,n})^2} | p,n \rangle \langle p,n | \right] \, .
\end{eqnarray}

Using the explicit expressions for $V^{(i)}$ and simplifying we obtain~\footnote{Notice that this expression coincides with Eq.(58)
of Ref.~\cite{Amore10b}, calculated in a different basis.}
\begin{eqnarray}
E_0^{(3)} &=& - \epsilon_{0,1}  \langle 0,1 | \sigma | 0,1\rangle^3
+\epsilon_{0,1}^3  \langle 0,1 | \sigma | 0,1\rangle \langle 0,1 | \sigma \hat{\Lambda} \sigma| 0,1\rangle  \nonumber \\
&+& 3 \epsilon_{0,1}^2  \langle 0,1 | \sigma | 0,1\rangle \langle 0,1 | \sigma \hat{\Omega} \sigma| 0,1\rangle
- \epsilon_{0,1}^3  \langle 0,1 | \sigma \hat{\Omega} \sigma \hat{\Omega} \sigma| 0,1\rangle \, .
\end{eqnarray}

For $\beta \rightarrow 0^+$ we have
\begin{eqnarray}
\lim_{\beta \rightarrow 0^+} E_0^{(3)} &=& \lim_{\beta \rightarrow 0^+} \epsilon_{0,1}^3 \left[
  \langle 0,1 | \sigma | 0,1\rangle \langle 0,1 | \sigma \hat{\Lambda} \sigma| 0,1\rangle
-  \langle 0,1 | \sigma \hat{\Omega} \sigma \hat{\Omega} \sigma| 0,1\rangle \right] \, .
\end{eqnarray}

Using the expressions for $G$ and $F$ in \ref{appa} we can obtain the explicit form of the
third order correction
\begin{eqnarray}
\lim_{\beta \rightarrow 0^+} E_0^{(3)} &=&
\frac{2 \pi^6}{b^9} \left( \int_{-\infty}^{\infty} dx_3\int_{-b/2}^{b/2} dy_3
 \cos^2\left(\frac{\pi  y_3}{b}\right) \sigma \left(x_3,y_3\right)\right)
\nonumber \\
&\times&  \int_{-\infty}^{\infty} dx_1\int_{-b/2}^{b/2} dy_1\int_{-\infty}^{\infty} dx_2\int_{-b/2}^{b/2} dy_2 \Big[
\left| x_1-x_2\right| \sigma \left(x_1,y_1\right) \sigma \left(x_2,y_2\right) \nonumber \\
& &  \times \cos ^2\left(\frac{\pi  y_1}{b}\right)  \cos^2\left(\frac{\pi  y_2}{b}\right)
- b \cos \left(\frac{\pi  y_1}{b}\right) \cos \left(\frac{\pi  y_2}{b}\right)
\nonumber \\
& & \times \sigma \left(x_1,y_1\right) \sigma\left(x_2,y_2\right)  {G}_2^{(0)}\left(x_1,y_1,x_2,y_2\right) \Big] \, .
\label{eq_third_order}
\end{eqnarray}

Notice that ${G}_2^{(0)}\left(x_1,y_1,x_2,y_2\right)$ is non analytic at $b=0$ and it
can be systematically approximated by
\begin{eqnarray}
\sum_{n=2}^\infty \frac{e^{-\frac{\pi \sqrt{n^2-1} \delta x}{b}}}{\pi \sqrt{n^2-1}}
&=& -\frac{e^{-\frac{\pi  \delta x}{b}}+\log \left(1-e^{-\frac{\pi  \delta x}{b}}\right)}{\pi}  \nonumber \\
&+&  \frac{-e^{-\frac{\pi  \delta x}{b}} (b+\pi  \delta x)+\pi
   \delta x {\rm Li}_2\left(e^{-\frac{\pi  \delta x}{b}}\right)+b
   {\rm Li}_3\left(e^{-\frac{\pi  \delta x}{b}}\right)}{2 \pi  b} +
 \dots
\end{eqnarray}

For $\delta x = |x_1-x_2| \gg b$ this expression decays exponentially while
for $\delta x \ll b$ the expression behaves as
\begin{eqnarray}
\sum_{n=2}^\infty \frac{e^{-\frac{\pi \sqrt{n^2-1} \delta x}{b}}}{\pi \sqrt{n^2-1}} &=&
-\frac{1}{\pi }+ \frac{3 \delta x}{2 b}-\frac{\log \left(\frac{\pi \delta x}{b}\right)}{\pi } +\frac{(\zeta (3)-1)}{2 \pi} +\frac{3 (\zeta (5)-1)}{8 \pi }
+ \dots
\end{eqnarray}

Therefore, ${G}_2^{(0)}\left(x_1,y_1,x_2,y_2\right)$ involves transversal modes and it correlates the perturbations of the density
in a region centered at the origin of size $\delta x  \ll b$. This behavior is somehow complementary to that of the first term which is
less sensitive to this region, due to the factor $|x_1 -x_2|$ appearing in the integral.

After inspecting eq.~(\ref{eq_third_order}) we discover that, whenever the second order contribution vanishes, the third order contribution vanishes
as well. To assess the presence of a bound state in this case, one should therefore calculate the further perturbative orders for the energy and
check the sign of the first non-vanishing term.


\section{Variational method}
\label{sec:VM}

In this section we want to derive the second order contribution obtained using perturbation theory, in
a much simpler way, using the variational theorem. In this case we use the formulation of the problem
corresponding to eq.~(\ref{eq:Helmholtz_ht}):
\begin{eqnarray}
-\nabla ^{2}\psi (x,y) &=& E \Sigma (x,y)\psi (x,y) \,  , \label{eq:Helmholtz}
\end{eqnarray}
where the solution obeys Dirichlet boundary conditions at the border
\begin{eqnarray}
\psi (x,\pm b/2) &=&0  ,
\end{eqnarray}
and decays at infinity
\begin{eqnarray}
\lim\limits_{|x|\rightarrow \infty }\psi (x,y) &=&0 \, .
\end{eqnarray}

The variational theorem provides an upper bound for the lowest eigenvalue
\begin{equation}
W = -\frac{\left\langle \varphi \right| \nabla ^{2}\left| \varphi \right\rangle }{\left\langle \varphi \right|
\Sigma \left| \varphi \right\rangle }\geq E_{0} \, . \label{eq:var_principle}
\end{equation}

We calculate $W$ using the trial function
\begin{equation}
\varphi =\sqrt{a}e^{-a|x|}\sqrt{\frac{2}{b}}\sin \left( \frac{n\pi [y+b/2]}{b}\right) \, ,  \label{eq:var_ansatz}
\end{equation}
where $a$ is a variational parameter. Working in the limit of weak inhomogeneities,
$\Sigma (x,y)=1+\sigma (x,y)$, with  $|\sigma (x,y)|\ll 1$, one obtains
the approximate value
\begin{equation}
a \approx \frac{\pi ^{2}\left\{ \int_{-\infty }^{\infty}\int_{-b/2}^{b/2}\sigma (x,y)
\cos {\left( \frac{\pi y}{b}\right) }^2 dy dx \right\} }{b^{3}}  \label{eq:a_var}
\end{equation}
and the approximate energy~\footnote{Notice that a bound state is present only if $a > 0$, implying  the condition
\begin{equation}
\int_{-\infty }^{\infty} \int_{-b/2}^{b/2} \sigma(x,y) \cos \left( \frac{\pi y}{b}\right)^2 dy dx  > 0 \, . \nonumber
\end{equation}
}
\begin{equation}
W\approx \frac{\pi ^{2}}{b^{2}}
-\frac{\pi ^{4}}{b^6}  \left( \int_{-\infty
}^{\infty }\int_{-b/2}^{b/2}\sigma {\left( x,y\right) }\left[ \cos \left( {%
\frac{\pi y}{b}}\right) \right] ^{2}\,dy\,dx\right)^2 \, ,
\label{eq:W_weak_lim}
\end{equation}
which coincides with the second order result obtained in the previous section.

\section{A solvable example}
\label{sec:example}

Consider an infinite heterogeneous waveguide of width $b$, parallel to the $x$ axis, and obeying Dirichlet boundary conditions on
$x = \pm b/2$. Let the density be
\begin{eqnarray}
\Sigma(x,y) = \left\{
\begin{array}{ccc}
1+\sigma  & ,   &  |x|< \delta/2 \, , \\
1 & , & |x|\geq\delta/2 \, .
\end{array}
\right.
\end{eqnarray}

The Helmholtz equation in this case is
\begin{eqnarray}
- \Delta \Psi(x,y) = E \Psi(x,y)
\end{eqnarray}
for $|x|>\delta/2$, and
\begin{eqnarray}
- \Delta \Psi(x,y) = E (1+\sigma)\Psi(x,y)
\end{eqnarray}
for $|x|<\delta/2$.

We look for a localized solution
\begin{eqnarray}
\Psi(x,y) = \sqrt{\frac{2}{b}} \sin \left(\frac{\pi {n_y} \left(\frac{b}{2}+y\right)}{b}\right) \times\left\{ \begin{array}{ccc}
a_1 e^{p_1 x}  & , & x <-\delta/2 \, , \\
a_2 cos ( p_2 x) & , & -\delta/2<x <\delta/2 \, , \\
a_3 e^{- p_1 x}  & , & x >\delta/2 \, , \\
\end{array} \right.
\end{eqnarray}
and impose the continuity of the solution and of its derivative at $x = \pm \delta/2$:
\begin{eqnarray}
a_1 &=& a_3 = a_2 \cos \left(\frac{d p_2}{2}\right) e^{\frac{1}{2} d p_2 \tan \left(\frac{d p_2}{2}\right)} \, , \\
p_1 &=& p_2 \tan \left(\frac{d p_2}{2}\right) \, .
\end{eqnarray}

Matching the solutions at the boundaries between the regions one obtains
\begin{eqnarray}
\frac{\pi ^2}{b^2}-p_2^2 \tan ^2\left(\frac{d p_2}{2}\right) = \frac{\frac{\pi ^2}{b^2}+p_2^2}{1+\sigma} \, ,
\end{eqnarray}
which provides a transcendental equation for $p_2$.

We can look for a solution to this equation analytically, taking the limit $\sigma \rightarrow 0^+$ and
expressing $p_2$ as a power series in $\sigma$:
\begin{eqnarray}
p_2^2 = \sum_{n=0}^\infty c_n \sigma^n  \, .
\end{eqnarray}

We find
\begin{eqnarray}
p_2^2 &=& \frac{\pi ^2 \sigma }{b^2}-\frac{\pi ^4 \delta^2 \sigma ^2}{4 b^4} + \frac{\pi ^4 \delta^2 \sigma ^3 \left(\pi ^2 \delta^2-3 b^2\right)}{12 b^6}
\nonumber \\
&+& \frac{\sigma ^4 \left(150 \pi ^6 b^2 \delta^4-23 \pi ^8 \delta^6\right)}{720 b^8}  + \frac{\pi ^6 \delta^4 \sigma ^5 \left(630 b^4-686 \pi ^2 b^2 \delta^2
+67 \pi ^4 \delta^4\right)}{5040 b^{10}}
+ \dots \nonumber 
\end{eqnarray}
and
\begin{eqnarray}
p_1 &=& \frac{\pi ^2 \delta \sigma }{2 b^2}-\frac{\pi ^4 \delta^3 \sigma ^2}{12 b^4}+\frac{\sigma ^3 \left(\pi ^6 \delta^5-5 \pi ^4 b^2
\delta^3\right)}{40 b^6} \nonumber \\
&+&\frac{\sigma ^4 \left(210 \pi ^6 b^2 \delta^5-23 \pi ^8 \delta^7\right)}{2520 b^8} +
\frac{\pi ^6 \delta^5 \sigma ^5 \left(1134 b^4-882 \pi ^2 b^2 \delta^2+67 \pi ^4 \delta^4\right)}{18144 b^{10}} + \dots \nonumber
\end{eqnarray}

The eigenvalue is
\begin{eqnarray}
E &=& \frac{\pi^2}{b^2}
- \frac{\pi ^4 \delta ^2 \sigma ^2}{4 b^4}
+ \frac{\pi ^6 \delta ^4 \sigma ^3}{12 b^6}
+ \frac{\sigma ^4 \left(90 \pi ^6 b^2 \delta ^4-23 \pi ^8 \delta^6\right)}{720 b^8} \nonumber \\
& & + \frac{\pi ^8 \delta ^6 \sigma ^5 \left(67 \pi ^2 \delta ^2-525 b^2\right)}{5040 b^{10}} + \dots
\end{eqnarray}

When we apply to this model the explicit formula obtained with perturbation theory (to second and third orders)
or with the variational theorem (to second order), we reproduce the coefficients of the formula above to
the corresponding orders.

\section{Conclusions}
\label{sec:conclusion}

We have derived the exact expressions for the energy of the fundamental mode of an infinite heterogeneous waveguide, in
two or more dimensions, to third order in the small parameter controlling the heterogeneity.
For the two-dimensional case our results also apply to an infinite, heterogeneous and locally deformed
waveguide. Our results show that, when the second order perturbative term does not vanish, a bound state is always present
when the heterogeneity corresponds to a small region of higher density; in the case in which the second order perturbative
term vanishes, the third order correction vanishes as well and it is not possible to determine whether a bound state exists.

\verb''\ack
The research of P.A and C.P.H. was supported by the Sistema Nacional de Investigadores (M\'exico).
F.M.F. thanks partial support from the Universidad Nacional de La Plata for the visit to the Universidad de Colima.

\appendix

\section{Green's function}
\label{appa}

We define the operator
\begin{eqnarray}
\hat{\Omega} \equiv  \left[{\sum_{n=2}^\infty}  \frac{1}{\epsilon_{0,n} - \epsilon_{0,1}}  | 0,n \rangle \langle 0,n |
+ {\sum_{n=1}^\infty} \int_0^\infty \frac{dp}{2\pi} \frac{1}{\epsilon_{p,n} - \epsilon_{0,1}} | p,n \rangle \langle p,n | \right] \, ,
\end{eqnarray}
which obeys the relation
\begin{eqnarray}
(\hat{H}_0 - \epsilon_{0,1}) \hat{\Omega} =  \left[{\sum_{n=2}^\infty}  | 0,n \rangle \langle 0,n |
+ {\sum_{n=1}^\infty} \int_0^\infty \frac{dp}{2\pi}  | p,n \rangle \langle p,n | \right] =
\hat{1} - | 0,1 \rangle \langle 0,1 | \, .
\end{eqnarray}

Equivalently,
\begin{eqnarray}
G(\mathbf{x}, \mathbf{x}') \equiv \langle \mathbf{x} | \hat{\Omega} | \mathbf{x}' \rangle
\end{eqnarray}
obeys the equation
\begin{eqnarray}
(\hat{H}_0 - \epsilon_{0,1})  G(\mathbf{x}, \mathbf{x}') = \langle \mathbf{x} | \hat{1} | \mathbf{x}' \rangle
-  \langle \mathbf{x} | 0,1 \rangle \langle 0,1 | \mathbf{x}' \rangle
= \delta (\mathbf{x} -\mathbf{x}' ) - \Psi_{01}(\mathbf{x}) \Psi_{01}(\mathbf{x}') \, .
\end{eqnarray}

Clearly
\begin{eqnarray}
\lim_{\beta \rightarrow 0^+} (\hat{H}_0 - \epsilon_{0,1})  G(\mathbf{x}, \mathbf{x}')
= \delta (\mathbf{x} -\mathbf{x}' )  \, .
\end{eqnarray}

Let us now work on the explicit expression of $ G(\mathbf{x}, \mathbf{x}')$:
\begin{eqnarray}
G(\mathbf{x}, \mathbf{x}') &=& \left[{\sum_{n=2}^\infty}  \frac{\phi_0(x) \phi_0(x') \psi_{n}(y) \psi_n(y')}{\epsilon_{0,n} - \epsilon_{0,1}}
+ {\sum_{n=1}^\infty} \int_0^\infty \frac{dp}{2\pi} \frac{\phi_p(x) \phi_p(x') \psi_{n}(y) \psi_n(y')}{\epsilon_{p,n} - \epsilon_{0,1}} \right]
\nonumber \\
&=& \int_0^\infty \frac{dp}{2\pi} \frac{\phi_p(x) \phi_p(x') \psi_{1}(y) \psi_1(y')}{\epsilon_{p,1} - \epsilon_{0,1}}  \nonumber \\
&+&\left[{\sum_{n=2}^\infty}  \frac{\phi_0(x) \phi_0(x') \psi_{n}(y) \psi_n(y')}{\epsilon_{0,n} - \epsilon_{0,1}}
+ {\sum_{n=2}^\infty} \int_0^\infty \frac{dp}{2\pi} \frac{\phi_p(x) \phi_p(x') \psi_{n}(y) \psi_n(y')}{\epsilon_{p,n} - \epsilon_{0,1}} \right]
\nonumber \\
&\equiv& G_{0}(\mathbf{x}, \mathbf{x}') + G_{1}(\mathbf{x}, \mathbf{x}')  + G_{2}(\mathbf{x}, \mathbf{x}') \, .
\end{eqnarray}
We observe that $\lim_{\beta \rightarrow 0^+} G_{0}(\mathbf{x}, \mathbf{x}') = \infty$,
and that $\lim_{\beta \rightarrow 0^+} G_{1}(\mathbf{x}, \mathbf{x}')$ exists:
\begin{eqnarray}
G_{0}(\mathbf{x}, \mathbf{x}') &=& \frac{1}{\beta} G_{0}^{(-1)}(\mathbf{x}, \mathbf{x}') + G_{0}^{(0)}(\mathbf{x}, \mathbf{x}')
+ \beta G_{0}^{(1)}(\mathbf{x}, \mathbf{x}')  + \dots \\
G_{1}(\mathbf{x}, \mathbf{x}') &=&  \beta G_{1}^{(1)}(\mathbf{x}, \mathbf{x}')  + \dots\\
G_{2}(\mathbf{x}, \mathbf{x}') &=&  G_{2}^{(0)}(\mathbf{x}, \mathbf{x}')
+ \beta G_{2}^{(1)}(\mathbf{x}, \mathbf{x}')  + \dots
\end{eqnarray}
The explicit expressions are
\begin{eqnarray}
G_{0}(\mathbf{x}, \mathbf{x}') &=&
\cos \left(\frac{\pi  y_1}{b}\right) \cos \left(\frac{\pi  y_2}{b}\right) \Big[ \frac{1}{2 b \beta } - \frac{\left| x_1\right| +2 \left|
   x_1-x_2\right| +\left| x_2\right| }{2 b} \nonumber \\
& & +O\left(\beta \right) \Big] \, ,
\end{eqnarray}
\begin{eqnarray}
G_{1}(\mathbf{x}, \mathbf{x}') &=& \frac{b \beta}{4\pi^2}  \Bigg( \cos \left(\frac{\pi  \left(y_1-y_2\right)}{b}\right)+\cos
   \left(\frac{\pi  \left(y_1+y_2\right)}{b}\right) \nonumber \\
&+&  2 i \Bigg(\log \left(-e^{-\frac{i \pi  \left(y_1-y_2\right)}{b}}\right)\sin \left(\frac{\pi  \left(y_1-y_2\right)}{b}\right)
+\log \left(e^{-\frac{i \pi  \left(y_1+y_2\right)}{b}}\right) \nonumber \\
& & \times \sin \left(\frac{\pi \left(y_1+y_2\right)}{b}\right) \Bigg) \Bigg) + O(\beta^2) \, ,
\end{eqnarray}
\begin{eqnarray}
G_{2}(\mathbf{x}, \mathbf{x}') &=& \sum_{n=2}^\infty \frac{e^{-\frac{\pi  \sqrt{n^2-1} \left(x_1+x_2\right)}{b}}}{\pi  \sqrt{n^2-1}}  \left(\theta
   \left(x_1-x_2\right) e^{\frac{2 \pi  \sqrt{n^2-1} x_2}{b}}+\theta
   \left(x_2-x_1\right) e^{\frac{2 \pi  \sqrt{n^2-1} x_1}{b}}\right) \nonumber \\
 & & \times  \sin\left(\frac{\pi  n \left(b+2 y_1\right)}{2 b}\right) \sin \left(\frac{\pi  n \left(b+2 y_2\right)}{2 b}\right)+ O(\beta) \, .
\end{eqnarray}

Similarly we define
\begin{eqnarray}
F(\mathbf{x}, \mathbf{x}') &\equiv& \langle \mathbf{x} | \hat{\Lambda} | \mathbf{x}' \rangle \nonumber \\
&=& \left[{\sum_{n=2}^\infty}  \frac{\phi_0(x) \phi_0(x') \psi_{n}(y) \psi_n(y')}{(\epsilon_{0,1} - \epsilon_{0,n})^2}
+ {\sum_{n=1}^\infty} \int_0^\infty \frac{dp}{2\pi} \frac{\phi_p(x) \phi_p(x') \psi_{n}(y) \psi_n(y')}{(\epsilon_{0,1} - \epsilon_{p,n})^2} \right]
\nonumber \\
&=& \int_0^\infty \frac{dp}{2\pi} \frac{\phi_p(x) \phi_p(x') \psi_{1}(y) \psi_1(y')}{(\epsilon_{0,1} - \epsilon_{p,1})^2}  \nonumber \\
&+&\left[{\sum_{n=2}^\infty}  \frac{\phi_0(x) \phi_0(x') \psi_{n}(y) \psi_n(y')}{(\epsilon_{0,1} - \epsilon_{0,n})^2}
+ {\sum_{n=2}^\infty} \int_0^\infty \frac{dp}{2\pi} \frac{\phi_p(x) \phi_p(x') \psi_{n}(y) \psi_n(y')}{(\epsilon_{0,1} - \epsilon_{p,n})^2} \right]
\nonumber \\
&\equiv& F_{0}(\mathbf{x}, \mathbf{x}') + F_{1}(\mathbf{x}, \mathbf{x}')  + F_{2}(\mathbf{x}, \mathbf{x}') \, .
\end{eqnarray}

For our present purposes it is sufficient to calculate $F_0$:
\begin{eqnarray}
F_{0}(\mathbf{x}, \mathbf{x}') &=& \cos \left(\frac{\pi  y_1}{b}\right)
\cos \left(\frac{\pi  y_2}{b}\right)
\Bigg(\frac{1}{8 b \beta ^3} - \frac{\left| x_1\right| + \left| x_2\right| }{8 b \beta^2} \nonumber \\
& & +\frac{4 \left| x_1\right|  \left| x_2\right| +16 x_1 x_2-6 \left(x_1^2+x_2^2\right)}{32 b \beta } +  O(\beta^0) \Bigg) .
\end{eqnarray}

We find it convenient to report some identities which are useful in the calculation of the third (and higher) orders,
\begin{eqnarray}
\left(\hat{H}_0  - E_{0}^{(0)} \right) | 0,1\rangle &=& 0 \, ,
\end{eqnarray}

\begin{eqnarray}
\left(\hat{H}_0  - E_{0}^{(0)} \right) \hat{\Omega} &=& \hat{\Omega} \left(\hat{H}_0  - E_{0}^{(0)} \right) \nonumber \\
&=& \sum_{n=2} |0,n\rangle \langle 0,n| + \sum_{n=1}^\infty \int_0^\infty \frac{dp}{2\pi}
|p,n\rangle\langle p,n| \equiv  \hat{P} \, ,
\end{eqnarray}
where $\hat{P}$ is a projection operator ($\hat{P}^2 = \hat{P}$) that can be used to express the completeness of the basis
in the form
\begin{eqnarray}
|0,1\rangle \langle 0,1 | + \hat{P} &=& \hat{1} \, .
\end{eqnarray}

In terms of this operator we also have
\begin{eqnarray}
\sigma \hat{P} \sigma &=& \sigma^2 - \sigma |0,1\rangle \langle 0,1 | \sigma  \, , \\
\sigma \hat{P} \left(\hat{H}_0  - E_{0}^{(0)} \right) &=&  \sigma \left(\hat{H}_0  - E_{0}^{(0)} \right) \, .
\end{eqnarray}

Finally
\begin{eqnarray}
\left(\hat{H}_0  - E_{0}^{(0)} \right) \hat{\Lambda} &=& \hat{\Omega} \, .
\end{eqnarray}

\section{Alternative perturbation theory}
\label{appb}

The purpose of this appendix is to obtain the main perturbation
equations in an alternative way. To this end we define
\begin{equation}
H=-\Delta -2\beta \delta (x) \, ,  \label{eq:H}
\end{equation}
where $\beta >0$. The problem becomes
\begin{equation}
H\psi _{0}=E_{0}(1+\lambda \sigma )\psi _{0} \, ,  \label{eq:H_psi}
\end{equation}
where $\lambda $ is a dummy perturbation parameter.

We assume that
\begin{equation}
H\left| n\right\rangle =\epsilon _{n}\left| n\right\rangle
,\;\epsilon _{0}<\epsilon _{1}\leq \epsilon _{2}\leq \ldots \, ,
\label{eq:H|n>}
\end{equation}
and that the lowest eigenvalue $\epsilon _{0}$ is non-degenerate.
In this case the label $n$ denotes a collection of quantum numbers
that describe all the degrees of motion of the problem. Then we
have
\begin{equation}
\left( H-\epsilon _{0}\right) \psi _{0}=\Delta E\psi _{0}+\lambda
E_{0}\sigma \psi _{0},\;\Delta E=E_{0}-\epsilon _{0} \, .
\label{eq:(H-e)_psi}
\end{equation}
If we apply $\left\langle 0\right| $ from the left and resort to
the intermediate normalization
\begin{equation}
\left\langle 0 \! \right. \left| \psi _{0}\right\rangle
=1,\;\left\langle 0 \! \right. \left| 0\right\rangle =1 \, ,
\label{eq:normalization}
\end{equation}
then equation (\ref{eq:(H-e)_psi}) reduces to
\begin{equation}
\Delta E=-\lambda E_{0}\left\langle 0\right| \sigma \left| \psi
_{0}\right\rangle \, .  \label{eq:Delta_E}
\end{equation}

We define the projection operator
\begin{equation}
P=1-\left| 0\right\rangle \left\langle 0\right| \, ,  \label{eq:P}
\end{equation}
that satisfies $[H,P]=0$, $P^{2}=P$ and $P\left| 0\right\rangle
=0$. It is also useful to define the operator
\begin{equation}
G=P\left( H-\epsilon _{0}\right) ^{-1}P \, ,  \label{eq:G}
\end{equation}
that enables us to rewrite equation (\ref{eq:(H-e)_psi}) as
\begin{equation}
P\psi _{0}=\Delta EG\psi _{0}+\lambda E_{0}G\sigma \psi _{0} \, .
\label{eq:P_psi}
\end{equation}
Note that
\begin{eqnarray}
\left( H-\epsilon _{0}\right) G^{k} &=&G^{k-1} \, ,\;k>1,  \nonumber \\
\left( H-\epsilon _{0}\right) G &=&P \, .  \label{eq:(H-e)G^k}
\end{eqnarray}
We obviously have
\begin{equation}
G^{k}=P\left( H_{0}-\epsilon _{0}\right) ^{k}P=\sum_{n \neq
0}\frac{\left| n\right\rangle \left\langle n\right| }{\left(
\epsilon _{n}-\epsilon _{0}\right) ^{k}} \, .
\end{equation}

In order to solve equations (\ref{eq:Delta_E}) and
(\ref{eq:P_psi}) we apply perturbation theory and expand
\begin{eqnarray}
\psi _{0} &=&\left| 0\right\rangle +\sum_{j=1}^{\infty }\psi
_{0}^{(j)}\lambda ^{j} \, ,  \nonumber \\
E_{0} &=&\epsilon _{0}+\sum_{j=1}^{\infty }E_{0}^{(j)}\lambda
^{j} \, ,
\end{eqnarray}
so that
\begin{equation}
\left\langle 0\right. \left| \psi _{0}^{(k)}\right\rangle
=0 \, ,\;k>0.
\end{equation}
Now equations (\ref{eq:Delta_E}) and (\ref{eq:P_psi}) become
\begin{equation}
E_{0}^{(k)}=-\sum_{j=0}^{k-1}E_{0}^{(j)}\left\langle 0\right|
\sigma \left| \psi _{0}^{k-j-1}\right\rangle ,\;k>0 \, ,
\end{equation}
and
\begin{equation}
\psi _{0}^{(k)}=\sum_{j=1}^{k}E_{0}^{(j)}G\psi
_{0}^{(k-j)}+\sum_{j=0}^{k-1}E_{0}^{(j)}G\sigma \psi
_{0}^{(k-j-1)} \, .
\end{equation}

For the first three orders we have: 

$k=1$
\begin{eqnarray}
E_{0}^{(1)} &=&-\epsilon _{0}\left\langle 0\right| \sigma \left|
0\right\rangle  \, ,  \nonumber \\
\psi _{0}^{(1)} &=&E_{0}^{(1)}G\left| 0\right\rangle +\epsilon
_{0}G\sigma \left| 0\right\rangle =\epsilon _{0}G\sigma \left|
0\right\rangle \, , \label{eq:k=1}
\end{eqnarray}

$k=2$
\begin{eqnarray}
E_{0}^{(2)} &=&-\epsilon _{0}^{2}\left\langle 0\right| \sigma
G\sigma \left| 0\right\rangle +\epsilon _{0}\left\langle 0\right|
\sigma \left|
0\right\rangle ^{2} \, ,  \nonumber \\
\psi _{0}^{(2)} &=&\epsilon _{0}E_{0}^{(1)}G^{2}\sigma \left|
0\right\rangle +\epsilon _{0}^{2}G\sigma G\sigma \left|
0\right\rangle +E_{0}^{(1)}G\sigma \left| 0\right\rangle \, ,
\label{eq:k=2}
\end{eqnarray}

$k=3$
\begin{eqnarray}
E_{0}^{(3)} &=&\epsilon _{0}^{3}\left\langle 0\right| \sigma
G^{2}\sigma \left| 0\right\rangle \left\langle 0\right| \sigma
\left| 0\right\rangle -\epsilon _{0}^{3}\left\langle 0\right|
\sigma G\sigma G\sigma \left| 0\right\rangle +3\epsilon
_{0}^{2}\left\langle 0\right| \sigma G\sigma \left| 0\right\rangle
\left\langle 0\right| \sigma \left| 0\right\rangle
\nonumber \\
&&-\epsilon _{0}\left\langle 0\right| \sigma \left| 0\right\rangle
^{3}. \label{eq:k=3}
\end{eqnarray}
These equations lead to the results derived in Section~\ref{sec:PT}. In particular, note that $G=\hat{\Omega}$
and $G^{2}=\hat{\Lambda}$.

\verb''\section*{References}
\verb''

\end{document}